\documentclass[journal]{IEEEtran}
\usepackage{draftwatermark}
\SetWatermarkText{DRAFT}
\SetWatermarkScale{2}
\SetWatermarkColor[gray]{0.9}
\usepackage{xcolor}
\usepackage{booktabs}
\usepackage{array}
\ifCLASSINFOpdf
\else
\fi
\usepackage{dblfloatfix}

\usepackage[most]{tcolorbox}
\usepackage{enumitem}

\definecolor{headercolor}{RGB}{51,51,51}
\definecolor{lightteal}{RGB}{230,245,245}

\newtcolorbox{findingsbox}[1]{
  enhanced,
  width=\columnwidth,
  colback=lightteal,
  coltitle=white,
  colframe=black,
  colbacktitle=headercolor,
  fonttitle=\normalsize\normalfont,
  title=#1,
  arc=1mm, 
  boxrule=1pt,
  top=5pt,
  bottom=5pt,
  left=5pt,
  right=5pt,
  fontupper=\small
}

\begin{document}

\title{Human-Centered Ambient and Wearable Sensing for Automated Monitoring in Dementia Care: A Scoping Review}

\author{Mason~Kadem,~\IEEEmembership{Member,~IEEE},
        Sarah~Masri,
        Anthea~Innes,
        and~Rong~Zheng,~\IEEEmembership{Senior~Member,~IEEE}
\thanks{M. Kadem and R. Zheng are with the Department of Computing and Software, McMaster University, Canada.}
\thanks{S. Masri is with the Department of Sociology, McMaster University, Canada.}
\thanks{A. Innes is with the Department of Health, Aging, and Society, McMaster University, Canada. and Centre for Rural Health Sciences, University of the Highlands and Islands, Scotland.}
\thanks{Manuscript submitted July, 2025;}}

\maketitle

\begin{abstract}
We conducted a scoping review to map the rapidly evolving landscape of wearable and ambient sensing technologies for monitoring people with dementia across home and institutional settings. We analyzed empirical sensing studies (2015-2025) to identify and inform future technical and human-centered design requirements. Five key implementation principles emerge: (1) human-centered design involving all stakeholders to augment rather than replace caregivers; (2) personalized, adaptable solutions that support autonomy across settings and severity levels instead of standardized approaches; (3) integration with existing workflows with adequate training and support; (4) proactive privacy and consent considerations, especially for ambient monitoring of residents and caregivers; and (5) cost-effective, ethical, equitable, scalable solutions with quantifiable outcomes. This paper identifies gaps, trends and opportunities for developing sensing systems that address the complex challenges, while enhancing automation and autonomy, in dementia care.
\end{abstract}


\IEEEpeerreviewmaketitle

\section{Introduction}

\IEEEPARstart{A}{s} of January 2025, the number of people with dementia worldwide is more than twice that of all cancers combined,  and is growing rapidly \cite{ADI2022, Armstrong2023}. While commonly viewed through a medical lens, dementia presents conditions beyond memory changes \cite{livingston2020dementia}, including progressive loss of executive function \cite{Arvanitakis2019}, language difficulties \cite{Kramer1996}, navigation problems~\cite{norberg2019sense}, and behavioural changes that alter how individuals perceive their world \cite{Feast_Orrell_Charlesworth_Melunsky_Poland_Moniz-Cook_2016}. Traditional clinical assessments provide only infrequent and inconsistent snapshots of cognitive and functional status, often missing subtle changes that signal important transitions in care needs. To mitigate these challenges, emerging automated monitoring through sensing technologies offers an opportunity to support both people with dementia and their care partners. Automated sensing technologies support people with dementia and caregivers by assessing cognitive changes and enhancing safety through continuous monitoring in both home and institutional settings. These technologies operate along a spectrum from wearable to ambient approaches, each with distinct advantages. Wearable sensors provide precise physiological tracking throughout daily activities, while ambient environmental sensors enable passive, unobtrusive monitoring without requiring user interaction, a critical advantage as cognitive abilities change.

Understanding the capabilities and limitations of these approaches is important because sensing technologies enhance autonomy by providing personalized support at the right time and in the right context\cite{rayyan-178408690}. While different monitoring contexts of home versus institutional care settings necessitate distinct design requirements, they can form a complementary monitoring continuum. Home monitoring solutions can inform and transition with the person into institutional settings, providing historical data and personalized baselines. Institutional monitoring practices can be adapted for home use, creating consistent support across changing living situations. This synergy between environments enables better care transitions and consistent monitoring approaches throughout the dementia journey. Home environments typically prioritize unobtrusive integration \cite{rayyan-164100656} and caregiver support, while institutional settings must balance privacy with safety monitoring across multiple residents \cite{rayyan-164101205}. Understanding these contextual differences and potential synergies is crucial for successful technology implementation across the dementia care continuum.


Current monitoring needs reflect diverse stakeholder perspectives. Administrators prioritize consistency and cost-effectiveness, clinicians focus on measurable outcomes and clinical validity, care partners need technology that reduces workload without adding technical burden \cite{covidtechpaper}, and patients require autonomy and privacy\cite{rayyan-164101012}. These perspectives often conflict, particularly regarding privacy versus safety trade-offs, highlighting the importance of inclusive design approaches. All stakeholders should participate in technology design as they often have varying concerns about monitoring approaches.

The existing literature on dementia monitoring technologies often create artificial divisions between human-centered and technical perspectives. Reviews in this field tend to focus narrowly on technology-centered aspects or soley on human-centered aspects. For instance, Lazar {\it et al}. \cite{rayyan-164100824} provided an insightful critical analysis of dementia and design in Human-Computer Interaction through a human-centered lens, yet their work largely overlooked the technical implementation challenges that are essential for translating concepts into viable monitoring solutions for care environments.

\begin{figure}[t]
    \centering
    \includegraphics[width=1\linewidth]{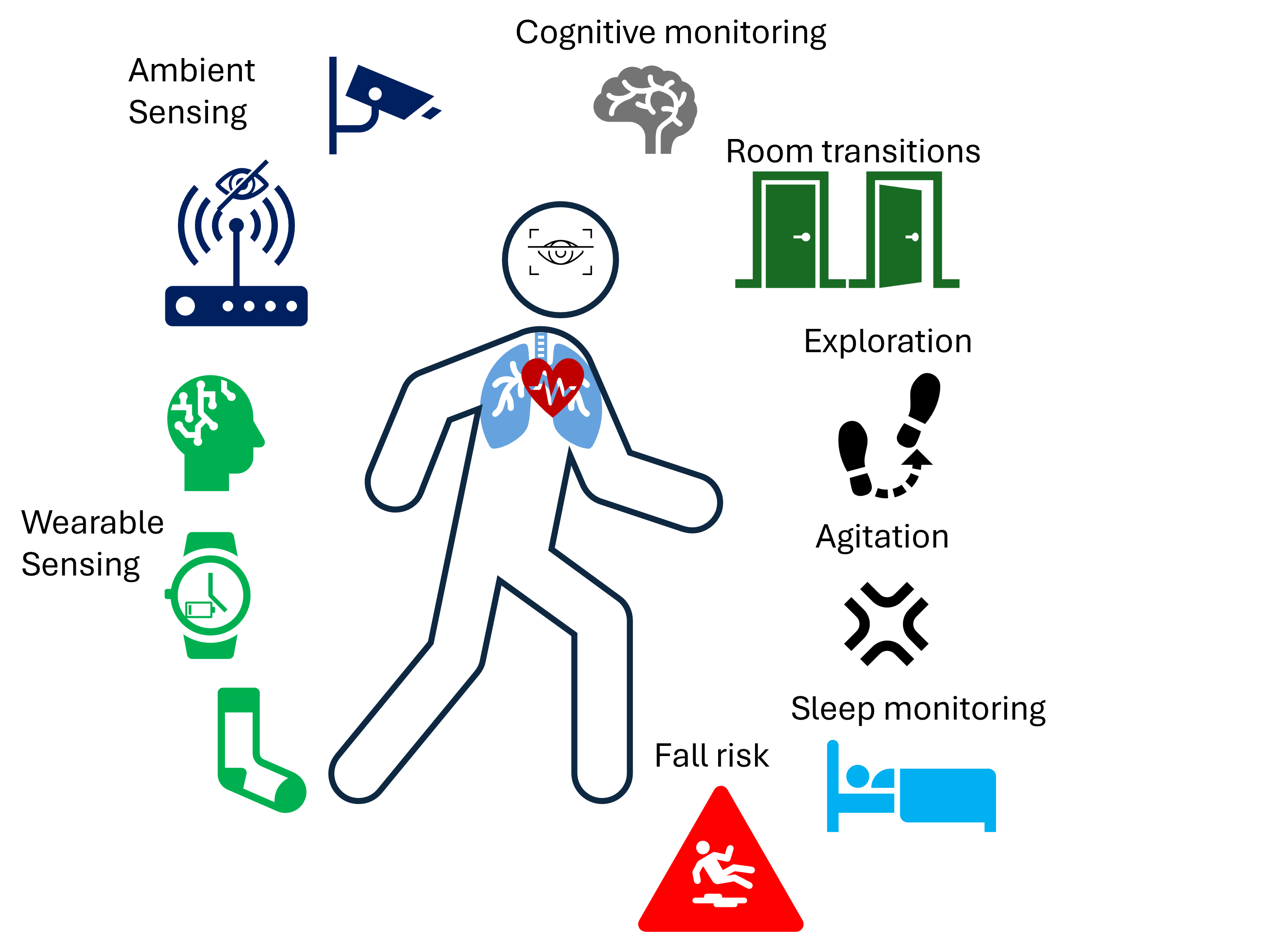}
    \caption{Schematic of ambient and wearable sensing use cases for dementia monitoring.}
    \label{fig:usecases}
\end{figure}

Existing reviews often restrict their scope to isolated sensing approaches. Zolfaghari {\it et al}. \cite{rayyan-164100660} focused exclusively on motion data for neurodegenerative disorders, providing valuable insights on gait analysis but neglecting other important monitoring domains such as sleep, activities of daily living, cognitive changes and behavioural symptoms. Salehi {\it et al}. \cite{Salehi2022} examined only wearable devices for dementia patients, missing the potential of ambient monitoring approaches that may be more suitable for those with advanced dementia who struggle with device compliance. Other reviews have restricted their focus to mild cognitive impairment \cite{rayyan-164101199} or non-dementia specific populations \cite{rayyan-164100995}, or single settings \cite{rayyan-164100652, rayyan-164100713} and do not examine implementation requirements across the care continuum and without incorporating multiple stakeholder perspectives. The fragmentation of knowledge makes it difficult for stakeholders to develop comprehensive monitoring strategies across the continuum of care.

Our analysis is guided by Value-Sensitive Design principles, ensuring technologies are evaluated for how they support human values in dementia care. Our review addresses these gaps by: 
\begin{enumerate}
\item  centering multidisciplinary human-centered design principles that position technology as adaptive support rather than caregiver replacement;
\item  identifying opportunities in wearable and ambient sensing across cognitive, safety, and behavioural monitoring;
\item  analyzing implementation requirements across home care and institutional settings, including the transitions between them;
\item  integrating stakeholder perspectives, ethical considerations, and adoption challenges. 
\end{enumerate}
By comparing wearable and ambient sensing approaches across different applications and settings, the integrated approach offers a practical evidence-based guidance for developing human-centered sensing systems for dementia care, bridges the critical gap between technical innovation and practical implementation, addressing the most common barriers to successful technology adoption in real-world care settings. Our review distinguishes itself by bridging these disciplinary divides, integrating both human-centered design principles and technical implementation considerations to identify gaps, trends and opportunities of dementia monitoring technologies.

\section{Methods}


To map the landscape of sensing technologies in dementia care, we conducted the scoping review following the Arksey and O'Malley framework \cite{Arksey2005}, enhanced by Levac {\it et al}. \cite{Levac2010} and the PRISMA-ScR guidelines \cite{Tricco2018}. The structured methodology employs five stages: (1) formalizing the research question, (2) identifying relevant studies, (3) study selection, (4) organizing the data, and (5) summarizing, analyzing and reporting results. We searched PubMed, IEEE Xplore, ACM Digital Library, PsychInfo, ProQuest, and Web of Science using the following comprehensive query structured around \textbf{population} (dementia-related terms), \textbf{concept} (sensing technologies), and \textbf{context} (care settings and implementation) dimensions:

\begin{quote}
\normalsize
(dementia OR alzheimer* OR ``cognitive impairment''
OR ``cognitive decline'' OR ``cognitive change'') AND
(sensor* OR wearable* OR radar OR ``ambient sensing'' OR ``ambient intelligence'' OR ``smart home'' OR
``remote monitoring'' OR ``Internet of Things'' OR IoT)
AND (``home care'' OR ``care home'' OR ``memory care''
OR ``nursing home'' OR ``aging-in-place'' OR ``assisted
living'' OR ``long-term care'') AND (adoption OR implementation OR barriers OR feasibility OR usability) NOT
``dementia diagnosis'' AND NOT ``risk prediction''
\end{quote}

\begin{figure}[ht]
\centering
\includegraphics[width=0.5\textwidth]{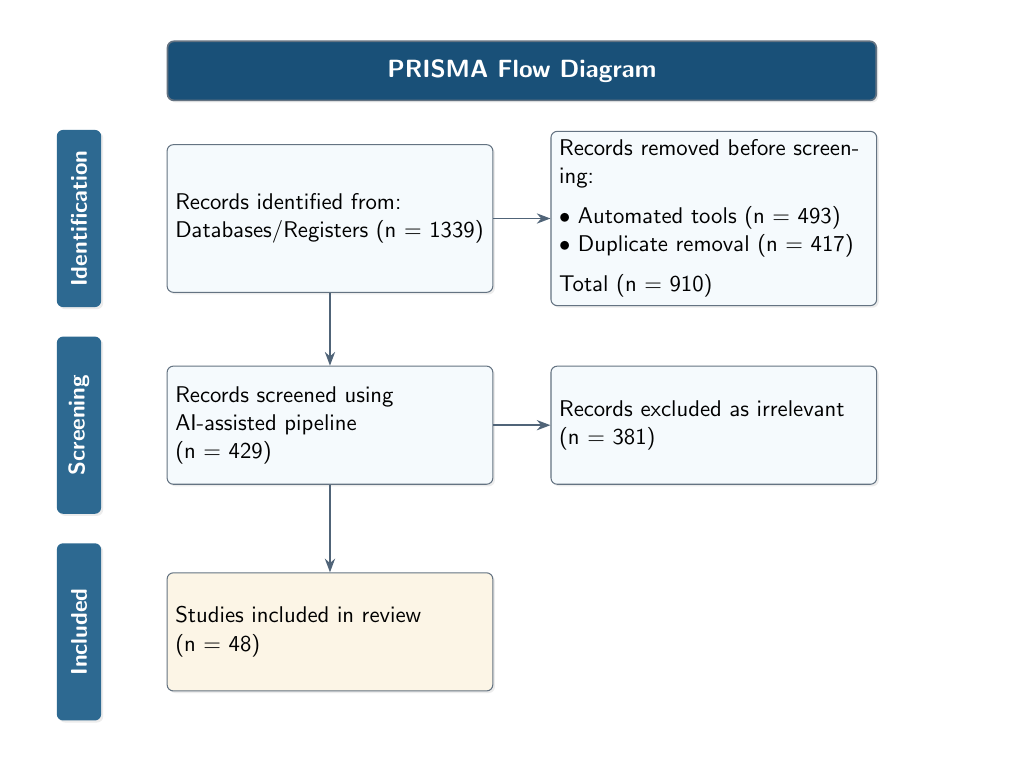}
\caption{PRISMA flow diagram showing the study selection process.}
\label{fig:prisma}
\end{figure}

Studies were included if they: (i) evaluated sensing technologies in home/memory care settings; (ii) involved people with dementia caregivers; (iii) addressed implementation outcomes; (iv) were peer-reviewed English publications (2015-2025); and (v) provided empirical implementation data. Studies were excluded as irrelevant if they: (i) focused solely on diagnosis/risk assessment; (ii) involved only non-cognitive impaired populations; (iii) examined non-sensing technologies; (iv) presented theoretical frameworks without implementation data; or (v) were reviews. Rayyan~\cite{ouzzani2016rayyan}, an AI-assisted screening tool, enhanced the literature selection process.

Out of 1339 initial records, 48 papers met our inclusion criteria after screening. 15 used wearable sensors, 18 used ambient sensing approaches, and 15 used both. 22 addressed home care environments, 15 institutional settings, and 11 spanned both settings. Methodologically, 29 papers presented technical development and validation, while 19 investigated user experiences and implementation practices. The balanced distribution across technology types, care settings, and research approaches enabled comprehensive analysis of the sensing technology landscape for dementia care, which we now examine in detail beginning with wearable sensor systems.


\section{Sensing Technology}

Sensing technologies used in dementia care generally fall into two categories: wearable and ambient sensing (Figure \ref{fig:usecases} and Table \ref{tab:wearable-ambient-comparison}). Each approach offers distinct advantages and limitations that affect their practical implementation and adoption. This section examines both approaches in detail, beginning with wearable sensor systems that directly contact the body, followed by ambient sensing systems that monitor without physical attachment to the human body.

\begin{table}[htbp]
\centering
\caption{Wearable vs. Ambient Sensing Tradeoffs}
\label{tab:wearable-ambient-comparison}
\footnotesize
\begin{tabular}{lcc}
\toprule
\textbf{Feature} & \textbf{Wearable} & \textbf{Ambient} \\
\midrule
Compliance & Required ↑ & Not needed ↓ \\
Privacy concern & Lower ↓ & Variable  \\
Best for stage & Early ← & Advanced → \\
Precision & High ↑ & Moderate ↓ \\
Setup complexity & Simple ↓ & Complex ↑ \\
Cost & High ↑ & Low ↓  \\
Scalability & Low ↓ & High ↑ \\
Portability & Mobile & Fixed \\
Multi-person & One person & Shared space \\
Maintenance & Daily ↑ & Minimal ↓ \\

\bottomrule
\end{tabular}
\end{table}

\subsection{Wearable Sensing}

We now examine wearable sensors for dementia care, exploring their underlying technical principles. Wearable sensors may employ various sensing modalities including accelerometry \cite{rayyan-164101141,rayyan-193570727, rayyan-193570620, rayyan-178408705}, gyroscopes, magnetometers, photoplethysmography, electrodermal activity, and temperature sensing. These technologies differ significantly in their sensing principles, data acquisition rates, and operational parameters.

Accelerometers, which measure acceleration relative to free fall, operate at variable sampling rates suitable for detecting human movement in healthcare applications, considerably lower than aerospace applications. The sampling rate impacts battery life which is important to consider for dementia monitoring. Accelerometers vary in sensitivities with lower sensitivity ranges being most appropriate for detecting subtle movements in daily activiies. 

Inertial Measurement Units (IMUs) combine accelerometers, gyroscopes, and sometimes magnetometers to provide comprehensive motion tracking. These multi-sensor packages enable more sophisticated analysis of movement quality beyond simple activity detection, allowing researchers to detect subtle gait abnormalities that may indicate cognitive changes. Phillips {\it et al}. \cite{rayyan-178408705} demonstrated that sensor placement significantly affects accuracy, with ankle mounted accelerometers showing substantially better agreement with observed step counts than wrist-worn devices in older adults with slower gait speeds typical of dementia.

Beyond movement tracking, physiological sensors capture additional dimensions of health status through autonomic nervous system activity. Photoplethysmography (PPG) sensors, typically operating at 25-1000~Hz, measure blood volume changes in the microvascular bed of tissue, enabling applications ranging from vital sign monitoring to cognitive assessment \cite{rayyan-164100844}. These sensors can detect heart rate, heart rate variability, and even estimate blood pressure when paired with advanced algorithms.  Electrodermal activity (EDA) sensors measure skin conductance changes at typical sampling rates of 4-32 Hz, providing insights into sympathetic nervous system activation associated with stress and agitation which are key concerns in dementia care. For example, EDA electrodes embedded in socks can monitor stress signals in people with dementia \cite{10914190}, enabling earlier intervention before agitation escalates. To shift towards less intrusive monitoring, these sensors are increasingly embedded directly in smart textiles like socks \cite{10914190}, shirts, and gloves \cite{rayyan-178408671}. Embedding sensors into everyday clothing helps overcome compliance challenges while maintaining the close body contact required for high-quality data acquisition. 

Recent innovations include both personalized approaches and smart textiles. Personalized solutions employ multiple sensor modalities. For example, Iaboni {\it et al}. \cite{Iaboni2022} combined wrist-worn accelerometers, passive infrared motion sensors, and pressure mats with personalized machine learning models. Unlike general-purpose activity trackers that apply universal thresholds, these systems recognizing that symptoms manifest differently across patients, enabling precise identification of changes that warrant intervention. 

While most wearable sensors focus on peripheral physiology, neural activity can also be monitored using electrodes placed on the scalp. Electroencephalography (EEG) has been used to detect patterns associated with falls~\cite{rayyan-164100814} and cognitive changes~\cite{Shim2022}, extending monitoring capabilities beyond autonomic responses. Integrated systems like BESI~\cite{rayyan-193570504} combine PPG, EDA, and skin temperature to detect early signs of agitation before they manifest visibly. Such approaches align with other physiological measures of cognitive load and effort, such as pupillometry, which has been shown to sensitively track cognitive processing demands even when they are not consciously perceived \cite{kadem2020pupil}. Webcam-based approaches offer potential for low-cost passive monitoring \cite{kadem2017webcam}, though validation in dementia populations is needed.

Advancing beyond individual sensors, the integration of these technologies into everyday wearable devices has significantly improved accessibility and acceptance to people with dementia. These advances in component-level design have enabled the development of systems that can monitor multiple physiological parameters simultaneously, providing a more comprehensive picture of a person's physical and emotional state than single-modality approaches. For example, smartwatches combine accelerometry, gyroscopes, and photoplethysmography to support multimodal monitoring \cite{ rayyan-178408670, rayyan-164101231, rayyan-178408672}. 

While wearable sensors offer advantages in direct physiological monitoring, they may face several barriers to adoption in dementia care. Compliance is a central challenge. Individuals may forget, remove, or resist devices due to cognitive impairment, discomfort, or perceived stigma. Technical constraints, including frequent charging and unreliable connectivity, limit usability. Additional concerns include skin irritation, safety risks, privacy, and poor integration with clinical workflows, all of which hinder long-term deployment.

\subsection{Ambient Sensing}

While wearable sensors offer direct physiological monitoring, ambient sensing technologies provide similar capabilities through fundamentally different operating principles, with each modality offering distinct advantages and limitations for dementia care monitoring. Ambient sensors monitor without physical attachment to the human body. These technologies include infrared motion sensors \cite{rayyan-178408635, rayyan-164101134, rayyan-178408677}, pressure sensors in beds, furniture, and floors \cite{rayyan-164101194, rayyan-193570818, rayyan-164101170}, computer vision based privacy-protected camera systems \cite{rayyan-164100657,  rayyan-206269304, rayyan-164101265}, LiDAR~\cite{rayyan-178408607} radar technologies \cite{rayyan-164100656, rayyan-164101208, rayyan-178408607, 10.1145/3589347}, acoustic sensors \cite{rayyan-164101224, rayyan-164100829}, environmental sensors \cite{rayyan-178408669, rayyan-178408661, rayyan-193570686}.

LiDAR (Light Detection and Ranging) technology, in contrast, employs pulsed laser light to measure distances to objects, typically operating at near-infrared wavelengths between 905-1550 nm. Indoor LiDAR systems for healthcare applications typically offer range resolutions of 1-5 cm with scanning rates between 5-20 Hz, sufficient for detecting human movements but presenting significant data processing challenges. Park {\it et al}. \cite{rayyan-178408607} developed an indoor LiDAR sensor specifically designed for institutional dementia care. The system combines low-power optoelectronics with neural processing to detect motion and falls with high precision. Unlike camera-based systems, that may raise privacy concerns, LiDAR technology provides detailed spatial information about movements without capturing identifiable images of residents. This approach addresses a critical balance between safety monitoring and privacy preservation, particularly in shared institutional environments.

Radar technology, particularly millimeter-wave (mmwWave) radar operating in the 30 - 300 GHz range, offers several unique capabilities. Unlike optical systems, radar can operate in darkness, through clothing, and is less susceptible to environmental conditions like dust or humidity. This technology is particularly valuable for monitoring residents during sleep without disruption. Zhang {\it et al}. \cite{10.1145/3589347} demonstrated a physiology-inspired approach to vital sign monitoring using mmWave radars that could extract respiratory and cardiac signals even with micro-level random body movement.

Recent advances in computer vision and privacy-by-design~\cite{rayyan-164100657} approaches have enabled systems that extract clinically relevant information without compromising privacy.
Edge computing, where all analysis is performed locally, prevents transmission of any sensitive video data, directly addressing stakeholder concerns about data security~\cite{rayyan-164100657} . The use of depth-only imaging eliminates visual identification while still enabling accurate activity recognition~\cite{rayyan-164101265}. . Processing only specific features needed for care while discarding identifying information further enhances privacy protection~\cite{rayyan-206269304}. Beyond activity recognition, camera-based systems can extract physiological parameters through remote photoplethysmography. These techniques detect subtle color changes in skin regions to measure heart rate and respiratory patterns without physical contact. When combined with the privacy-preserving approaches validated by [45] and the low-resolution techniques demonstrated by [47], these systems can provide comprehensive health monitoring while maintaining privacy.

Pressure sensing arrays embedded in floors, furniture, and beds represent another important category of ambient sensing. These systems typically use resistive, capacitive, or piezoelectric sensors with spatial resolutions ranging from 5-50 cm and pressure sensitivities capable of distinguishing between walking, sitting, and lying postures. 

While not traditionally considered ambient sensing due to its requirement for subject interaction, Radio Frequency Identification (RFID) offers a complementary monitoring approach. Unlike the ambient systems described above, RFID requires individuals or objects to carry tags that are activated by nearby readers,  enabling both location tracking of individuals and monitoring of interactions with tagged objects (typically operating in frequency ranges of 125 kHz to 960 MHz). Passive RFID tags, powered by the electromagnetic energy transmitted by the reader, offer advantages of low cost, indefinite operational life, and small form factor make them ideal for large-scale deployment throughout living environments. Mibu and Shinomiya \cite{rayyan-178408634} applied this technology by embedding passive RFID sensor tags in everyday objects throughout the home environment, demonstrating accurate detection of activities such as meal consumption and trash disposal by monitoring object interactions. This approach exemplifies how ambient technology can be integrated into the environment rather than requiring residents to interact with or wear devices, potentially increasing end-user acceptance by balancing between comprehensive monitoring and preserving autonomy.

Complementing individual sensor types, environmental sensor networks integrate multiple modalities to create comprehensive monitoring infrastructures. Environmental sensor networks combine various sensor types including passive infrared motion detectors, temperature/humidity sensors, light sensors, and door contact switches to create a comprehensive monitoring infrastructure. These systems generate multi-dimensional datasets that, when analyzed with advanced algorithms, can reveal patterns of daily living and detect deviations that may indicate cognitive or functional changes \cite{rayyan-164100708}.

Each of these ambient sensing modalities presents different privacy implications, installation requirements, and data characteristics. The selection of appropriate technologies must be guided by the specific monitoring needs, environmental constraints, and ethical considerations relevant to each care setting. Having examined the technical foundations of both wearable and ambient sensing approaches, we now turn to their practical applications in dementia care, exploring how these technologies address specific monitoring needs across cognitive, safety, and behavioural domains.

\section{Applications in Dementia Care}

The sensing technologies described in the previous section enable several critical applications in dementia care, each addressing different aspects of the condition and supporting specific care requirements (Table ~\ref{tab:clinical-sensing-matrix}). In this section, we analyze how both wearable and ambient approaches are applied to two primary domains: cognitive change monitoring and safety/behaviour monitoring. For each domain, we examine the distinct advantages and limitations of different sensing approaches.

\begin{table*}[htbp]
\centering
\caption{Clinical Domains and Sensing Technologies in Dementia Care. Summary of relevant markers, sensor modalities, and use cases organized by clinical domain with representative examples from reviewed studies.}
\label{tab:clinical-sensing-matrix}
\footnotesize
\setlength{\tabcolsep}{6pt}
\renewcommand{\arraystretch}{1.3}
\begin{tabular}{l>{\raggedright\arraybackslash}m{2.5cm}>{\raggedright\arraybackslash}m{3.2cm}>{\raggedright\arraybackslash}m{3.2cm}>{\raggedright\arraybackslash}m{3.2cm}@{}}
\toprule
\textbf{Domain} & \textbf{Markers} & \textbf{Sensors} & \textbf{Assessment} & \textbf{Care} \\
\midrule

\textbf{Fall Risk} & 
Gait parameters, turning frequency, balance metrics, acceleration patterns & 
Wearable IMUs, pressure-sensitive floors, LiDAR, mmWave radar, computer vision & 
Fall risk prediction, gait abnormality detection, balance assessment, wandering identification & 
Fall detection/prevention, mobility intervention, safety monitoring, wandering management \\

\midrule

\textbf{Sleep} & 
Sleep fragmentation, bed exit events, cardiac and motion patterns, circadian disruptions & 
Bed pressure sensors, infrared sensors, radar technology, wearable sleep trackers, PPG, passive IR motion sensors & 
Sleep variability, cognitive changes, sleep disorder identification, nighttime wandering detection & 
Sleep intervention planning, nighttime monitoring, bed exit alerts, sleep hygiene support \\

\midrule

\textbf{Cognition} & 
Routine activity deviations, repetitive behaviours, heart rate variability, activity patterns & 
Environmental sensor networks, PPG, activity trackers, RFID systems, eye-tracking & 
Cognitive change detection, executive function, activity pattern analysis & 
Activity reminders, cognitive support, task feedback, trajectory monitoring \\

\midrule

\textbf{Behaviour} & 
Agitation signatures, electrodermal activity, movement changes, physiological stress indicators & 
Accelerometers, electrodermal sensors, integrated multimodal systems, environmental sensors, acoustic sensors & 
Agitation detection, stress level monitoring, behaviour identification, emotional state assessment & 
Early agitation intervention, behavioural management, workload reduction, stress-triggered support \\

\bottomrule
\end{tabular}
\end{table*}

\subsection{Cognitive Change Monitoring}

Dementia progressively impairs cognitive abilities, affecting memory, executive function, and daily task performance. Despite advances in detecting cognitive changes, including cognitive testing, biomarkers (e.g., p-tau181~\cite{Grande2025}), brain imaging (CT, MRI, PET)~\cite{Karikari2020}, and genetic screening \cite{Koriath2020}, cognitive changes remain underdiagnosed in clinical settings, lacks standardization in research practice, and leaves many affected individuals and their caregivers with unmet needs~\cite{livingston2020dementia}. Cognitive assessments suffer from inconsistency and rater reliability, while brain imaging, genetics and biomarkers remain too expensive and invasive for routine monitoring.

Wearable and ambient sensing technologies provide promising alternatives through continuous, passive monitoring of multiple digital markers. Subtle changes in sensory and motor function may precede clinical dementia symptoms by 10 to 15 years, offering an early detection window that traditional neuropsychological tests may miss~\cite{Digitalb25:online}. Gait analysis can detect characteristic changes in walking patterns (speed, stride length) that begin deteriorating a decade before cognitive symptoms appear, measurable through sensors with high accuracy~\cite{Digitalb25:online}. Fine motor control assessment through finger tapping and tracing tests can identify early indicators of neurological changes, while eye tracking technology utilizing standard device cameras can continuously monitor pupillary response and eye movement patterns associated with cognitive decline.

Heart rate variability, easily measured through common wearable devices~\cite{Rykov2024}, serves as a particularly valuable indicator by reflecting autonomic nervous system function, which deteriorates early in dementia's progression due to cholinergic system disruption. Additionally, sleep monitoring can detect fragmentation patterns and circadian rhythm disruptions characteristic of cognitive decline, while passive tracking of social engagement, mobility patterns, and daily activities provides insight into functional changes that may signal underlying neurobiological deterioration. These digital biomarkers collectively offer unprecedented opportunities for early, non invasive, and continuous monitoring of cognitive health. Motion data analysis enables detection of subtle behavioral changes such as agitation, wandering, and sleep disturbances that may indicate early cognitive changes. Cognitive neuroscience researchers have developed models describing abnormal activity routines that may indicate early symptoms of mild cognitive impairment. 

Recent work has shown that non-invasive and accessible markers can reliably detect cognitive changes remotely and provide decision support\cite{kadem2023xgboost}, while multimodal approaches combining neuroimaging, genetics and clinical data have demonstrated strong predictive performance for cognitive changes~\cite{kadem2023interpretable}. Despite this potential, few sensor systems have been designed to detect cognitive changes remotely.

We now examine how these principles are implemented through both wearable and ambient sensing approaches for cognitive monitoring.

\subsubsection{Wearable Sensing for Cognitive Change Monitoring}

Building on their technical capabilities described earlier, wearable sensors support cognitive monitoring through several mechanisms. For example, wrist-worn sensors continuously track activity patterns to identify subtle changes in routine activities indicative of cognitive changes \cite{rayyan-193570838, rayyan-193570792}, supplementing traditional clinical evaluation \cite{rayyan-193570838}. Wu {\it et al}. \cite{rayyan-164100708} used an In-Home Multi-Sensor system (i.e., motion, bed, and depth sensors) to track behavioural trajectories over years, demonstrating the potential for early health change detection in elderly residents. Using data from over 2,400 older adults, researchers demonstrated that accelerometer-derived features such as sleep parameters, activity summaries, and light exposure, along with user-reported data like age and education can detect cognitive changes~\cite{Sakal2024}.

The monitoring of physiological correlates of cognition has shown promising results \cite{rayyan-164101150, rayyan-164100844}. Gwak {\it et al}. \cite{rayyan-164100844} demonstrated that photoplethysmography signals can distinguish between healthy controls and individuals with mild cognitive impairment with classification  accuracy of 0.90, outperforming traditional neuropsychological measures alone. Their research with 62 older adults (average age 72 years) demonstrated that parameters derived from photoplethysmography measurements (i.e., Maximum PPG amplitude and interval of PPG peaks)  could improve the detection of cognitive changes. Using longitudinal data from a 10-week clinical trial, researchers examined the association between physiological signals collected via wearable devices and cognitive performance in individuals with mild cognitive impairment. Physiological features, particularly heart rate variability, showed strong associations with executive function scores. By combining these features with demographic data, models were developed to predict cognitive performance across four domains: global cognition, memory, executive function, and processing speed. The findings support the feasibility of using wearable-derived physiological data as a proxy for cognitive function, enabling passive, real-time assessment and continuous monitoring in ambulatory settings. Leveraging sleep stage electroencephalogram(EEG)-based brain index may also provide an objective assessment of cognitive states \cite{Associat24:online}. 

A qualitative questionnaire gathered needs and requirements for assistive technology from 45 end-users including healthcare professionals, caregivers, and people with mild cognitive impairment \cite{rayyan-164101190}. Adoption of technology is influenced by ease of use, personal need, and privacy concerns. Caregivers, often stressed by constant monitoring, noted that technology could reduce patrols and ease concerns about sleep, behaviour, vital signs, and medication. MCI participants expressed interest in using technology to support cognitive health. Mild cognitive impairment participants emphasized challenges with sleep, and given the impact of dementia on sleep remote monitoring solutions should include sleep tracking. Recent work has demonstrated that interpretable wearable-based approaches can achieve competitive sleep staging performance while maintaining transparency \cite{kadem2025sleep}.

While wearable approaches offer direct physiological measures for cognitive monitoring, ambient sensing systems provide complementary monitoring capabilities without requiring user compliance, an important advantage as cognitive abilities decline.

\subsubsection{Ambient Sensing for Cognitive Change Monitoring}

Ambient sensing systems are particularly effective for long-term cognitive monitoring (although understudied) as they operate through passive observation without requiring residents to actively engage with technology \cite{rayyan-164101228, rayyan-178408677, rayyan-178408691}. Environmental sensor networks combine motion sensors, temperature sensors, and door switches to create predictive models by identifying deviations from established activity patterns \cite{rayyan-164101228, rayyan-206269243, rayyan-206269218, rayyan-178408634}. A study by Riboni {\it et al}. \cite{rayyan-164101246} utilized fixed passive infrared sensors and door contact sensors in residential settings to detect subtle deviations in daily routines that might signal cognitive changes. Their approach analyzed sensor activation sequences to identify unusual temporal patterns in activity and incorporated a unique combination of acoustic signal processing with three distinct text mining methodologies to detect verbal events. The researchers distinguished their work by focusing on fine-grained recognition of abnormal behaviours for early detection of mild cognitive impairment. Motion and door sensors can be integrated with applications like a digital memory notebook to support functional independence by providing real-time feedback on completed tasks and timely reminders for pending ones, addressing memory lapses as they occur. This smart-home integration is especially valuable for individuals with cognitive impairments, as it reduces the need for manual programming and extends their ability to live independently \cite{rayyan-193570792}.

Another study employed passive infrared sensors embedded throughout the home to monitor behavioural deviations, an early indicator of cognitive changes \cite{rayyan-164101228}. By tracking patterns of forgetfulness, such as repeatedly entering rooms without completing intended tasks, the system can flag changes in behaviour. For instance, if a person forgets the reason for entering a room, like going from the kitchen to the bedroom and back, it can signal cognitive impairment. Such monitoring can assist in early detection and provide valuable insights into behavioural changes associated with cognitive changes. While some empirical studies focus on mild cognitive impairment, the approaches could be extended to assess people further along in their dementia journey.

Bed sensors equipped with pressure-sensitive components can detect sleep state variations that may predict cognitive changes \cite{rayyan-164101194, rayyan-164101170, rayyan-193570818}. Chen {\it et al}. \cite{rayyan-178408714} demonstrated that bed pressure sensors could detect sleep state variability with sufficient accuracy to predict mild cognitive impairment in community-dwelling seniors with over 70\% recall using just two months of data \cite{rayyan-164101194}. This suggests non-invasive sleep monitoring can serve as an effective early screening tool for cognitive changes. 

Emerging approaches such as privacy-preserving video monitoring show potential for building cognitive assessment tools. For example, one study deployed a filtered video-based monitoring system in an assistive technology center, revealing user acceptance when privacy is respected \cite{rayyan-164100657}. A study collected data from seven single-resident apartments over two weeks demonstrated that low-resolution depth cameras could recognize daily activities such as standing, sitting, TV interaction, and inactivity with high accuracy \cite{rayyan-164101265}, supporting the potential of these systems to monitor \cite{rayyan-164101265}, enabling prediction of activities or actions of people with dementia to support their functions. 

Beyond cognitive monitoring, sensing technologies play a critical role in addressing the safety challenges and behaviuoral symptoms that accompany dementia progression. The following section examines how these technologies support safety monitoring and behavioural management.

\subsection{Safety and Behaviour Monitoring}

As dementia progresses, safety and behavioural challenges intensify (e.g., increased fall risk, exploration, agitation), creating risks for individuals and cognitive burden for caregivers. This section examines how wearable and ambient sensing technologies address safety management and behavioural symptom detection through continuous automated monitoring. 

\subsubsection{Wearable Sensing for Safety and Behaviour Monitoring}

Wearable technologies detect falls~\cite{rayyan-164100814, rayyan-164100806, rayyan-164101079}, recognize pre-fall movement patterns using EEG systems~\cite{rayyan-164100814}; identify wandering using body-worn inertial sensors~\cite{rayyan-193570620, rayyan-164101018}; capture physiological signatures of agitation and stress~\cite{rayyan-164101141, rayyan-193570727, rayyan-164101283}, monitor bed exits~\cite{rayyan-178408635}, assess gait and mobility issues~\cite{rayyan-164100765}; and track sleep patterns \cite{rayyan-178408680, rayyan-193570818}. 

The physiological and motion signatures of agitation can be detected through various wearable approaches. A study across two nursing care facilities for a period of 4 weeks employed multi-axis accelerometers attached to both wrists and ankles of 17 nursing home residents with moderate to severe dementia to monitor motion patterns associated with agitation \cite{rayyan-193570727}. When staff members were debriefed qualitatively, they indicated that incorporating the technology and observation platform into regular nursing home procedures was feasible. The results demonstrated that their system could accurately detect sensor activities and showed that the problem of noncompliance seemed manageable, with only 11\% of recordings affected. 

A wearable GPS-based tracking system was developed to support caregivers in monitoring Alzheimer’s patients in nursing homes \cite{rayyan-164101205}. The system included a transceiver for real-time location tracking and notifications when patients wandered beyond predefined boundaries. The study emphasized the importance of continuous monitoring in high-risk populations and demonstrated the feasibility of adopting wearable sensors to enhance safety and reduce caregiver burden in institutional settings.

A multimodal monitoring system integrated continuous data streams from PIR motion sensors, sleep mats, and appliance sensors placed in participants’ homes~\cite{Bafaloukou2025}. The PIR sensors captured activity patterns, indoor light levels, and temperature, while sleep mats provided heart rate, respiratory rate, and nighttime movement data. External environmental context was incorporated using weather and light data sourced from the Visual Crossing Weather API. Their retrospective analysis modeled weekly patterns to better understand factors contributing to agitation.A low respiratory rate and high indoor artificial illuminance were identified as key indicators of agitation, suggesting that both physiological and environmental factors play critical roles in the development of agitation.  Importantly, the model approach therein provided actionable, interpretable insights that caregivers could use to mitigate the risk of agitation. 

Kamil {\it et al}. \cite{rayyan-193570620} used strategically placed body-worn inertial sensors to track clinically relevant movement patterns. Inertial sensors positioned on the body identify wandering behaviour by quantifying turning frequency and direction changes, a precursor to wandering events. The investigators provided initial evidence supporting the feasibility of continuous monitoring in dementia patients using wearable sensors and suggested further research to determine whether objective measurements of turning behaviours collected through inertial sensors could reliably identify wandering behaviour. They found that wanderers make more frequent, quicker turns compared to non-wanderers, consistent with pacing behaviour.

The integration of multiple sensing modalities enhances the capacity to monitor complex behaviours. Davidoff {\it et al}. \cite{rayyan-178408596} integrated physiological sensors (PPG, EDA), environmental sensors (light, sound, temperature), and wearable accelerometers to quantify agitation in dementia by analyzing multiple data streams together. Their analysis identified three key themes regarding technology integration in care: the placement of technology within care contexts, the substitution of care activities with technological solutions, and the potential displacement of traditional care practices by technology, highlighting both opportunities and implementation challenges. Their study also included contextual, physiological, and psychological parameters in a study with multiple participants.

While multi-agent systems \cite{rayyan-164100795} and improved hierarchical algorithms \cite{rayyan-164101111} improve detection of complex activity patterns, computational approaches from other medical domains could further enhance physiological monitoring in dementia care. Advanced computational methods combining hemodynamic modelling, medical imaging and machine learning for cardiovascular assessment \cite{Kadem2023} offer transferable approaches for analyzing sensor-derived physiological data to detect subtle changes associated with agitation or distress.

\subsubsection{Ambient Sensing for Safety and Behaviour Monitoring}

Ambient sensing systems have shown strong potential to support older adults who wish to live independently despite cognitive or physical change. In a 490-day longitudinal case study,an ambient sensing system was deployed to monitor the daily activities of a 90-year-old woman with Alzheimer’s disease, capturing data on sleep patterns, mobility, outings, hygiene, and cooking behaviour~\cite{rayyan-178408677}. These trends were compared with information gathered by a clinical nurse through regular interviews. The system revealed changes in behaviour over time, including some not yet identified by the nurse. For example, the system detected shifts in routine following meal preparation. Most of the behavior patterns captured by the system were consistent with clinical observations, confirming validity. These findings highlight the value of ambient sensing systems in identifying subtle behavioural changes that may otherwise go unnoticed. 

Radar sensing provide continuous monitoring without wearable devices, addressing compliance challenges \cite{rayyan-164100656, rayyan-164101208}, while better preserving privacy than video based systems. Some systems  monitor nighttime movement and wondering \cite{rayyan-178408635, rayyan-164100812}. For example, The eNightLog system integrates two complementary sensing modalities: an infrared 3D time-of-flight sensor (Kinect V2 sensor with RGBV camera concealed) and an impulse-radio ultra-wideband sensor to detect actions (e.g., sitting up or getting out of bed) and monitors respiration rate (for presence/orientation), respectively. Together, these sensors enable precise detection of bed exits and wandering behaviour without requiring wearables or physical contact, and support low false alarm rates, addressing common issues in fall detection systems such as alarm fatigue and excessive caregiver workload\cite{rayyan-178408635}. A recent study introduced a cloud-based, ambient system for real-time activity and gait monitoring in home environments using mmWave radar and deep learning models \cite{10012054}. Trained on range-Doppler maps from real-world in-home data, the system achieved 93\% accuracy for trained users and 86\% for new users. It passively tracks activity levels, washroom use, sleep patterns, and gait parameters, while preserving privacy by eliminating the need for wearable devices.

Another study deployed a low-cost, low-power indoor LiDAR-based monitoring system designed for fall detection in long-term care facilities for Alzheimer’s patients \cite{rayyan-178408607}. The system uses a CMOS-based optoelectronic front-end with on-chip  photodiodes and a neural processing unit for motion analysis. By leveraging approximate computing and network sparsity, the sensor operates efficiently while maintaining real-time fall detection capabilities. The ambient LiDAR sensor offers a promising, non-intrusive solution for improving safety and monitoring.
 
Acoustic sensors enable passive tracking of gait biomarkers without intrusive monitoring. A study with 10 community-dwelling older adults, showed that gait analysis could be performed in residential settings using only footstep sounds~\cite{rayyan-164101224}. Bland-Altman Analysis evaluated the agreement between parameter values from three different monitoring systems, confirming that their acoustic approach was reliable and comparable to more established methods. They found the gait parameters derived from their acoustics-only system exhibit relative standard errors of only 0.71\% and 0.58\% against IMU and video-based system, respectively. 

Computer vision systems using cameras augmented by privacy-preserving algorithms can assess gait parameters and predict falls non-intrusively \cite{rayyan-164101248,rayyan-164101265, rayyan-178408677, rayyan-193570686}. Adeli {\it et al}. \cite{rayyan-164101248} developed a computer vision-based monitoring system to predict fall likelihood within a 4-week timeframe. Their study with 54 older dementia patients demonstrated that their comprehensive system achieved an area under the ROC curve of 76.2\%, whereas the best model that excluded ambient gait features reached only 56.2\%, highlighting the essential contribution of ambient monitoring to accurate fall prediction. Future research will focus on externally validating these findings to prepare for implementation in long-term care facilities.

Low-resolution depth cameras have also demonstrated high accuracy in recognizing specific activities in real homes \cite{rayyan-164101265}. Separately, location-based environmental monitoring using ambient sensors offers additional context-aware support. Schinle {\it et al}. \cite{rayyan-178408720} developed an approach to support ambulatory care using passive infrared motion detectors and door-contact sensors designed to infer wake times, bedtimes and night activity including anomalies related to dementia. Their research addressed the challenge of extracting meaningful parameters from limited sensor data histories for individuals living alone who cannot manage wearable devices or provide feedback about their daily activities. They specifically started with day-night rhythm and night-time activity as relevant parameters, since these parameters are associated with dementia and beneficial for implementing care.

Analyzing movement and gait patterns using sensor data from a smart floor that correlate with fall risk~\cite{rayyan-164100765}. This study explored a machine learning approach leveraging smart floor pressure sensors to assess fall risk in older adults by analyzing foot pressure data during everyday activities. Using center of pressure values, the system extracted time and frequency domain features related to balance. These features were used to train models to predict Tinetti scores (a clinical measure of gait and balance). The system demonstrated potential for continuous, unobtrusive monitoring and early detection of fall risk, supporting wider adoption of sensor-based technologies in eldercare environments. Beyond floor-based systems, wearable approaches include a notable EEG-based system using genetic programming to classify falls from brainwave patterns, achieving 89\% accuracy by extracting wavelet features and applying Principal Component Analysis~\cite{rayyan-164100814}. This neural monitoring approach represents an innovative complement to traditional movement-based fall detection, potentially identifying neurological precursors to fall events. 

A multi-sensor ambient monitoring system was installed in the homes of older adults with mild cognitive impairment and Alzheimer’s disease to enable continuous, unobtrusive assessment of cognitive and daily functioning \cite{rayyan-164101134}. Over 4–12 months, participants receiving tailored interventions based on ambient sensor data showed improved cognitive outcomes, sleep quality, and activity levels. The study supports the feasibility and clinical value of long-term adoption of ambient sensor systems in home settings for cognitive care. 

A study leveraged ambient sensing for monitoring and guiding the behaviour of individuals with cognitive changes in home environments \cite{rayyan-164100693}. Using sensors to track patient location and activity, the system employed probabilistic model checking to predict behaviours and detect abnormalities. Daily routines were modeled as discrete-time Markov chains and analyzed against expected patterns to identify deviations in activity timing or frequency. The approach demonstrated the feasibility of real-time, sensor-based behaviour analysis and guidance, supporting long-term adoption of WSNs for enhancing safety, independence, and personalized care in populations with cognitive changes.

As illustrated above, each sensing approach offers distinct trade-offs between measurement precision, compliance requirements, privacy, and implementation complexity that must be carefully considered based on individual needs and care contexts. By offering continuous, objective insights, these systems can  fill gaps left by human observation, making them a valuable tool in resource-constrained healthcare environments. Ambient sensors provide safety and behavioural support through multiple applications, highlight the potential of sensing technologies in dementia care. But their real-world success depends not only on technical capability, but also on how well they align with human needs and care contexts. The following section outlines key human-centered design principles that support effective and ethical implementation.

\section{Human-Centered Design Requirements for Dementia Care}
\normalsize

When surveying sensing technology systems for dementia care, our analysis identified five critical human-centered design requirements that determine successful implementation and adoption. These include: (1) technical capabilities that function across multi-user environments and diverse physical settings, (2) addressing detection accuracy and reliability challenges unique to people living with dementia, (3) managing integration complexity across platforms and care settings, (4) embedding ethical principles and autonomy considerations into core design processes, and (5) balancing diverse stakeholder perspectives to ensure technologies deliver measurable end-user care outcomes. The interplay between these factors creates a complex implementation landscape that must be navigated thoughtfully to develop effective sensing solutions.

\subsection{Technical Design Requirements}

Technical and human factors impact adoption of sensing technologies in several key areas. 


\subsubsection{Multi-user environments}

A fundamental challenge involves distinguishing between multiple residents in shared living environments \cite{rayyan-164100995} and tracking multiple patients simultaneously in institutional settings \cite{rayyan-164101205}. Proposed solutions include differentiating individuals based on movement patterns while preserving social connections (e.g., recognizing visitors) and while maintaining privacy \cite{rayyan-164100656}. Tiersen {\it et al}. \cite{rayyan-178408669} employed user-centered design to create systems for households where people with dementia live with family members, demonstrating the importance of technologies that function in complex social environments. 



\subsubsection{Physical Environment Adaptation}

Systems must function effectively across different care settings, particularly during transitions between home care and memory care. Technologies must adapt to diverse physical environments. Li {\it et al}. (2025) \cite{rayyan-164100656} demonstrated advantages of radar approaches for fall detection in challenging environments such as bathrooms or poorly lit areas where traditional video systems fail. Abedi {\it et al}. \cite{rayyan-164101208} achieved high accuracy for contactless in-bed detection using a low-cost, low-resolution radar, showing how non-contact technologies can function in intimate settings while preserving privacy.

\subsubsection{Detection Accuracy and Reliability}

Current sensing technologies face important accuracy and reliability limitations in dementia care. Phillips {\it et al}. \cite{rayyan-178408705} showed that gait speed impacts activity monitoring accuracy, with most devices underestimating true steps in individuals with slower gait speeds typical of dementia. Location and orientation dependencies affect monitoring accuracy as well~\cite{rayyan-164100656}. Tewell {\it et al}. \cite{rayyan-193570686} identified limitations in monitoring meaningful activities beyond basic activities of daily living using small low-cost devices in a smart home. Their work revealed the need for more comprehensive monitoring approaches that capture a wider range of meaningful activities that contribute to quality of life, not just basic function. 
Gathercole {\it et al}. \cite{rayyan-164101119} found that assistive technology did not increase the time participants could continue living independently compared to a basic pendant alarm. Long-term data collection spanning years is needed to establish reliable baselines and detect meaningful changes \cite{rayyan-164100708}. Systems require both high sensitivity to detect critical events (like bed exits and wandering) and high specificity to minimize false alarms \cite{rayyan-178408635}. 

\subsubsection{Integration complexity}

The integration of multiple sensing modalities presents both technical and practical challenges, particularly in resource-constrained care environments. The current lack of interoperability standards means that data from different sensing systems often remain siloed, preventing the comprehensive understanding that could emerge from combined analysis. Research on the integration of multiple sensor modalities for monitoring agitation shows that while integrated systems provide more accurate detection,  implementation complexity presents barriers to adoption \cite{rayyan-178408596}. Loreti {\it et al}. \cite{rayyan-193570639} documented implementation challenges for complex event processing in assisted living environments, particularly regarding integration of sensor networks into existing homes. The field would benefit from standardized approaches to data integration that can work seamlessly across different sensing modalities and platforms. Research on wearable devices for assessing function in dementia found concerns about device complexity and battery life impact willingness to adopt technology \cite{rayyan-178408672}. Expected value, user comfort, need for information, and design considerations influence technology acceptance \cite{rayyan-178408671}. 

Research involving focus groups and interviews on sensor systems for stress detection with people with dementia, their families, and professional caregivers emphasized the need for easy-to-use, personalized systems, with educated stakeholders and clear guidelines for use~\cite{rayyan-178408658}. Adam {\it et al}. \cite{rayyan-178408725} highlighted the importance of addressing stakeholder concerns about washability, cost, safety, and potential replacement of human contact.

\subsection{Ethical Considerations}

Ethical considerations fundamentally shape successful implementation of supportive technologies. 

\subsubsection{Privacy}

The literature articulate nuanced views about technology's role in supporting residents and caregivers \cite{rayyan-164100657}. Hall {\it et al}. \cite{rayyan-178408690} moved beyond the simplistic "safety versus autonomy" dichotomy through a qualitative exploration of ethics in dementia care. Their findings revealed additional ethical dimensions often overlooked in previous research, including impacts on workforce dynamics when care professionals themselves are monitored. Staff reported concerns about surveillance of their work practices, which could affect morale, recruitment, and retention. The study also identified equity issues, as not all residents had equal access to beneficial technologies. These insights demonstrate how ethical implementation must consider the broader social and organizational context beyond individual resident concerns. Moreover, that their exists conflicting perspectives on privacy and safety between care recipients and caregivers, necessitating flexible approaches that accommodate varied perspectives

Privacy must be considered a core design requirement rather than an afterthought that risks limiting data collection. McNeill {\it et al}. \cite{rayyan-164101012} identified distinct reasons why older adults value privacy, including desires for autonomy, maintaining positive social identity, and challenging traditional social norms. Practical approaches to balancing safety and privacy concerns \cite{rayyan-164100656}, include customizable privacy settings and clear communication about data usage. But different stakeholders perceive the privacy-safety tradeoff differently \cite{rayyan-164100657, rayyan-178408690, rayyan-164100659}, necessitating flexible approaches that accommodate varied perspectives.

\subsubsection{Consent processes}

Ambient monitoring technologies present specific ethical considerations around consent \cite{rayyan-164100658}. Unlike wearable sensors where consent is typically implied (an assumption that may not hold for individuals with dementia), Ambient sensors enable monitoring without physical engagement, increasing the risk of non-consensual data collection. When implementing technologies with cognitively impaired individuals, consent should be viewed as an ongoing process rather than a one-time event~\cite{rayyan-164101119}.

\subsubsection{Autonomy-enhancing features}

Co-designing systems with people with dementia, families and other relevant stakeholders ensures technologies align with diverse values and preferences \cite{rayyan-178408690}. Several studies highlight how technology can amplify existing capabilities and support continued autonomy. Research provides frameworks for implementing systems that recognize and enhance individual strengths, preferences, and self-determination \cite{rayyan-178408690}. Adoption and continued use are low when technologies are perceived as reinforcing aging stigma or reminding older adults of negative aspects of aging \cite{rayyan-164100664}. Their work emphasizes designing technologies that celebrate diverse abilities rather than highlighting limitations.

Many older adults appreciate receiving personalized insights from sensor technology to support continued self-direction and health management \cite{rayyan-164100713}, valuing technologies that enhance their ability to make informed decisions and maintain control over daily routines. Innovative approaches to implementing monitoring systems include subtle environmental feedback that maintains autonomy rather than traditional alerting systems. Finally,technologies should grow and evolve alongside users, continually adapting to honor changing expressions of identity, values, and preferences \cite{rayyan-164100695}.

\subsection{Balancing Stakeholder Perspectives}

Multiple stakeholder perspectives influence technology adoption and implementation. Sensing technology supports the sustainability of entire care networks (i.e., long-term viability, resilience, and balance) rather than simply monitoring the person with dementia in isolation. Monitoring technology enables the expansion of informal care networks, allowing care duties to be distributed among more people. This redistribution alleviates the constant burden felt by primary caregivers. According to a study involving 14 family members~\cite{rayyan-178408691}, the benefits of this redistribution were perceived to outweigh privacy concerns. 

Research shows stakeholder preferences regarding decision-making responsibility change over the course of cognitive changes \cite{rayyan-164100655, Kademmhuman}, with control gradually shifting from the person with dementia to their caregivers. Studies assessing ambient assisted living support through questionnaires applied to healthcare stakeholders \cite{rayyan-164100669}, find strong support for implementation but identifying the need for careful consideration of individual profiles.

Caregiver acceptance of technology directly impacts quality of care and quality of life for people with dementia  \cite{rayyan-164100780}, with analysis of nursing care activity-related stress levels showing that technologies reducing caregiver burden and workload through automated monitoring are more likely to be adopted when they address staff needs and operational workflows. Involving people with dementia and their caregivers in the design process improves technology acceptance and ongoing use \cite{rayyan-178408672}.

Advanced supportive technologies must be tailored to specific care settings. In home environments\cite{rayyan-164100656, rayyan-164100653}, practical challenges include logistical challenges of retrofitting sensing systems in diverse residential settings~\cite{rayyan-178408839}, cultural views on home care and emphasizing the need for designs that adapt to aging~\cite{rayyan-164100656}. Smart aging care systems raise security concerns distinct from general smart \cite{rayyan-164100662}, necessitating clear protocols for data governance. The SAFE House project \cite{rayyan-178408839} found that technologies require clear protocols for data management and access to maintain privacy while providing safety benefits. 

Long-term evaluations of remote activity monitoring have shown potential to delay adverse health events~\cite{rayyan-178408664}, with ambient systems supporting prolonged independent living \cite{rayyan-193570613}, showing high levels of acceptability and utility among users \cite{rayyan-193570583}. 

In institutional settings, different design considerations emerge. Technologies must integrate with physical and operational infrastructure~\cite{rayyan-164101141}, align with staff workflows, and enhance rather than disrupt existing care relationships \cite{rayyan-164100780}.

\subsection{Measurable End-user Care Outcomes}

Massi \cite{rayyan-178408682} reported that responsive ambient sensing in nursing homes reduced falls by 71\% and prevented wandering, supporting continued mobility and exploration. Beyond clinical outcomes, the implementation demonstrated economic benefits (e.g., reduction in caregiver costs, more occupied beds, additional caregiver hiring). Supportive technologies transform care relationships, creating more meaningful interactions by reducing task-focused interactions and creating space for genuine connection \cite{rayyan-164100780}. Care partners reported more satisfaction and deeper relationships when routine checks were handled by technology, allowing them to focus on relational aspects of care.

Multi-sensor monitoring systems have demonstrated ability to detect subtle health changes weeks before they become clinically apparent, enabling early interventions that prevent hospitalization\cite{rayyan-164100708}. Monitoring technologies reduce caregiver workload and stress levels by automating routine checks, prioritizing care needs, and providing advance warning of potential issues \cite{rayyan-164100780}. Monitoring for challenging behaviours can improve care efficiency while simultaneously enhancing quality of life through less restrictive care approaches \cite{rayyan-164101283}, and technology-enabled personalization can address specific challenges like sleep disturbances in dementia care \cite{rayyan-164101170}.

\begin{figure}[!htbp]
\begin{findingsbox}{Recommendations for developing dementia-specific sensing technologies}
\label{tab:summary}
\begin{itemize}[leftmargin=10pt, topsep=0pt, itemsep=5pt, font=\small]
\item \textbf{Human-centered design involving all stakeholders is essential.} Technologies developed without input from people with dementia, caregivers, and clinicians consistently fail adoption regardless of technical sophistication~\cite{rayyan-178408672, rayyan-178408658}.

\item \textbf{Multi-person environments require accurate individual identification and minimal false alarms.} Multiple studies identified challenges distinguishing between residents in shared settings \cite{rayyan-164100995, rayyan-164100656}. Scalable systems must maintain high sensitivity and specificity to reduce caregiver fatigue \cite{rayyan-178408635}.

\item \textbf{Personalization over standardization addresses dementia's heterogeneity.}Given the wide variation in symptom presentation documented across studies, flexible multimodal approaches \cite{rayyan-164100693, rayyan-164100792,kadem2023xgboost, kadem2023interpretable} enable earlier detection and personalized decision support.

\item \textbf{Workflow integration is critical for implementation success.} Studies consistently report that technologies fail due to inadequate training and poor integration with existing care practices \cite{rayyan-193570639, rayyan-178408596}. Systems must complement established workflows with cross-setting continuity.

\item \textbf{Proactive ethical design must address privacy, consent, and workforce concerns.} The literature reveals that ambient technologies raise unique questions about consent for cognitively impaired individuals \cite{rayyan-164100658} and staff surveillance concerns \cite{rayyan-178408690}.

\item \textbf{Cost-effectiveness and demonstrated longitudinal outcomes drive adoption.} Validated outcomes like 71\% fall reduction \cite{rayyan-178408682} and reduced caregiver workload \cite{rayyan-164100780, rayyan-178408691} provide the evidence base needed to justify implementation investment. 
\end{itemize}
\end{findingsbox}
\end{figure}

\section{Discussion and Future Directions}

Beyond technical considerations, human-centered design remains a major factor in whether sensing systems succeed in real-world care settings. Our scoping review represents the most comprehensive synthesis to date of sensing technologies specifically designed and evaluated for dementia care. We synthesized findings from 48 empirical studies specific to dementia care that employed sensing technologies, with 15 using wearable sensors, 18 using ambient approaches, and 15 using both methodologies.

The literature demonstrates distinct advantages and limitations of wearable versus ambient sensing approaches in dementia care. Wearable systems provide precise physiological data but face significant adoption barriers. Multiple studies report adherence rates between 50-70\% in dementia populations \cite{rayyan-178408670, rayyan-193570727, Iaboni2022}, with additional challenges including device discomfort, charging requirements, battery limitations, skin irritation, and risk of loss. Even when worn, accuracy is compromised in people with slower gait speeds typical of dementia \cite{rayyan-178408705}.

Ambient technologies offer compelling advantages by eliminating compliance requirements entirely. As cognitive impairment advances, the ability to remember to wear, charge, or operate wearable devices diminishes. Ambient systems provide consistent monitoring regardless of disease progression, enabling longitudinal tracking from early to advanced stages.

Radar-based technologies \cite{rayyan-164100656, rayyan-164101208, rayyan-178408607} demonstrate particular promise for dementia care. The reviewed studies indicate these systems can function in challenging environments like bathrooms, operate regardless of lighting conditions, detect micro-level movements \cite{10.1145/3589347}, and preserve privacy better than camera-based systems. As noted in Section V, stakeholders express stronger acceptance of technologies that cannot capture identifiable images \cite{rayyan-178408690}.

Despite their advantages, ambient sensing technology faces integration challenges in real-world care environments. Multiple studies \cite{rayyan-193570639, rayyan-178408596} highlight the complexity of implementing sensor networks in existing infrastructure. The current lack of interoperability standards means that data from different sensing systems often remain siloed, preventing the comprehensive understanding that could emerge from combined analysis. Addressing these implementation barriers requires a systematic approach to technology development. 

From an IoT systems perspective, several architectural considerations emerge from the reviewed literature. Edge processing enables privacy-preserving analysis by keeping sensitive data local rather than transmitting to cloud servers \cite{rayyan-164100657, rayyan-178408607}. Data fusion across multiple sensor modalities improves detection accuracy~\cite{rayyan-178408596, rayyan-164100708} but requires standardized interfaces and communication protocols that are currently lacking. Scalability remains a concern, as most validated systems involved fewer than 20 participants. Institutional deployment serving dozens of residents simultaneously presents unresolved technical challenges. Future IoT architectures for dementia care should prioritize interoperability standards that enable seamless integration of wearable and ambient sensors across care settings.

A gap in current research is support for transitions between care settings. Transitions between home and institutional care create significant stress for people with dementia and their caregivers \cite{rayyan-164100801}. Ambient sensing technologies could potentially ease these transitions by providing continuity of information across settings without requiring the person with dementia to learn new devices or procedures during an already challenging period. 

Future research should focus on developing systems that function effectively across the continuum of care, from independent living to assisted living to memory care. Such systems would need to adapt to different environmental contexts, care protocols, and stakeholder needs while maintaining consistent core functionality and data continuity. 


\section{Limitations}

This review has several limitations. Our coverage of empirical sensing studies may have excluded valuable theoretical frameworks or non-sensing interventions that could inform implementation. The rapid evolution of sensing technologies means our 2015-2025 timeframe may miss recent commercial developments. Heterogeneity in study designs across the included papers limits direct quantitative comparisons. Most reviewed studies were small-scale feasibility trials conducted over weeks to months in high-income countries with predominantly white participants. This restricts generalizability across cultures, socioeconomic contexts, and dementia subtypes. Publication bias likely favors positive findings, and our English-language restriction may have excluded relevant international research. Finally, inconsistent validation methodologies across studies and the scarcity of head-to-head comparisons between sensing approaches limit our ability to make definitive recommendations about optimal technologies for specific use cases.

\section{Implications}

The findings from this review have practical implications for multiple stakeholder groups. Researchers should prioritize multi-year longitudinal studies, cross-setting validation protocols, and systematic examination of algorithmic fairness. Clinical teams considering technology adoption should ensure adequate staff training and workflow integration planning before deployment, as these factors predict success more reliably than technical specifications alone. Healthcare institutions would benefit from developing standardized technology assessment protocols that incorporate the human-centered principles identified in this review. For technology developers, interoperability and demonstrated outcomes in caregiver burden reduction should take priority alongside technical performance metrics. Policymakers have an opportunity to accelerate progress by establishing evaluation frameworks and reimbursement pathways that incentivize both innovation and equitable access across care settings.

\section{Conclusion}

This scoping review mapped both wearable and ambient sensing technologies for dementia care in home and institutionalized settings while integrating technical and human-centered perspectives across empirical studies. Beyond technical considerations, human-centered design remains a determining factor in whether sensing systems succeed in real-world care settings. Six implementation principles emerged from this analysis: (1) human-centered design involving all stakeholders to augment rather than replace caregivers, (2) accurate individual identification with minimal false alarms in multi-person environments, (3) personalized and adaptable solutions over standardized approaches, (4) integration with existing workflows alongside adequate training and support, (5) proactive attention to privacy and consent considerations, and (6) cost-effective and scalable solutions with demonstrated outcomes. These principles provide a framework for developing monitoring technologies that genuinely support people with dementia and their care partners. As sensing technologies continue to advance, maintaining focus on these human-centered requirements will be essential for translating technical capabilities into meaningful improvements in dementia care.

\bibliographystyle{IEEEtran}
\bibliography{refs}
\ifCLASSOPTIONcaptionsoff
  \newpage
\fi

\end{document}